\documentclass[aps,twocolumn,preprintnumbers,floats]{revtex4}\def\@cite#1#2{\textsuperscript{[{#1\if@tempswa , #2\fi}]}}
\usepackage{graphicx}
\usepackage{amsmath}
\usepackage{amsfonts}
\usepackage{amssymb}
\usepackage{color}
\usepackage{subfigure}
\usepackage{epsfig}
\usepackage{morefloats}
\usepackage{multirow}
\usepackage{graphicx,booktabs}
\usepackage{mathrsfs}
\usepackage{txfonts}
\usepackage{indentfirst}
\usepackage{graphicx,booktabs}
\usepackage{pifont}
\usepackage{bbding}
\usepackage{longtable,lscape}
\usepackage[colorlinks, citecolor=blue,anchorcolor=red,menucolor=red, linkcolor=red,filecolor=red,urlcolor=blue,frenchlinks=red]{hyperref}

\begin{document}

\title{  Dynamical study of $T_{ss}$ systems at a chiral quark model}

\author{Jiazheng Ji}
\email[E-mail: ]{3078203935@qq.com}
\affiliation{Department of Physics, Yancheng Institute of Technology, Yancheng 224051, P. R. China}

\author{Yuheng Xing}
\email[E-mail: ]{phyxyh@ycit.edu.cn}
\affiliation{Department of Physics, Yancheng Institute of Technology, Yancheng 224051, P. R. China}

\author{Xinxing Wu}
\email[E-mail: ]{wuxinxing@ycit.edu.cn}
\affiliation{Department of Physics, Yancheng Institute of Technology, Yancheng 224051, P. R. China}

\author{Ning Xu}
\email[E-mail: ]{xuning79530@126.com}
\affiliation{Department of Physics, Yancheng Institute of Technology, Yancheng 224051, P. R. China}

\author{Yue Tan}
\email[E-mail: ]{tanyue@ycit.edu.cn (Corresponding author)}
\affiliation{Department of Physics, Yancheng Institute of Technology, Yancheng 224051, P. R. China}

%
%\date{\today}

\begin{abstract}
Since the discovery of $T_{cc}$ by LHCb, there has been considerable interest in $T_{cc}$ and its heavy-flavor partners. However, the study of its strange partner $T_{ss}$ has been largely overlooked. Within the framework of the chiral quark model, we conducted a systematic study of the bound states of $T_{ss}$ utilizing the Gaussian Expansion Method. Considering all physical channels with $01^{+}$, including molecular and diquark structures. Our calculations revealed that upon considering the coupling between diquarks and molecular states, we identified a deep bound state with a bounding energy of 60 MeV, primarily composed of $K K^{*}$. Using the $^3P_0$ model, we calculated the decay width of $K^{*}$ within the $KK^{*}$ bound state, which is approximated as the decay width of the bound state in the $T_{ss}$ system.  The results indicate that due to the effect of binding energy, the decay width of $K^{*}$ in $KK^{*}$ is approximately $3$ MeV smaller than that of $K^{*}$ in vacuum. Additionally, resonance state calculations were performed. Utilizing the real-scaling method, we searched for possible resonance states in the $T_{ss}$ sysytem. Due to the strong attraction in the $[K^{*}]_8[K^{*}]_8$ configuration,  four resonance states were found in the vicinity of $2.2$-$2.8$ GeV, predominantly featuring hidden-color structures, and their decay widths are all less than $10$ MeV.  We strongly recommend experimental efforts to search for the resonance states in the $T_{ss}$ system  predicted by our calculations.
\end{abstract}

%\pacs{}

\maketitle

\section{Introduction}
In 1964, Gell-Mann~\cite{Gell-Mann:1964ewy} and Zweig~\cite{Zweig:1964jf} proposed the conventional quark model, suggesting that mesons and baryons are composed of $q\bar{q}$ and $qqq$, respectively. At that time, by utilizing this model, most of the hadrons were explained well.
Over the past two decades, however, there has been an influx of new experimental data unveiling properties of hadronic states that challenge the predictions of the conventional quark model, such as $X(3872)$~\cite{Belle:2003nnu}, $Y(4626)$~\cite{Belle:2019qoi}, $Z_{c}(3900)$~\cite{BESIII:2013ris} and so on. These exotic hadronic states provide a valuable platform for a deeper understanding of the properties of low-lying Quantum Chromodynamics (QCD).

In 2021, LHCb Collaboration reported a new state $T_{cc}^{+}(3875)$ in the $D^{0}D^{0}\pi$ mass distribution with a mass $M=3875.1+\delta m_{exp}$, where $\delta m_{exp}=-273\pm61\pm5^{+11}_{-14}$ keV, and a width $\Gamma=410\pm165\pm43^{+18}_{-38}$ keV~\cite{LHCb:2021auc,LHCb:2021vvq}. The very narrow width indicates that it is a long-living exotic state. Its spin-parity quantum number is $J^{P}=1^{+}$.
The minimal quark content of $T_{cc}^{+}(3875)$ is $cc\bar{u}\bar{d}$, which means that,
unlike the three hidden-charm tetraquark states mentioned above, there can be no annihilation for such a component.
These interesting properties have attracted widespread theoretical attentions and researches
~\cite{Collins:2024sfi,Ikeda:2013vwa,Junnarkar:2018twb,Du:2021zzh,Meng:2021jnw,Tan:2024pqs,
Wang:2022clw,Lu:2020rog,Meng:2020knc,Chen:2021vhg,Ebert:2007rn,Feijoo:2021ppq,Yan:2021wdl,Dong:2021bvy,Huang:2021urd,
Fleming:2021wmk,Braaten:2022elw,Hu:2021gdg,Chen:2021cfl,Deng:2021gnb,Wang:2022jop,Ling:2021bir,Ke:2021rxd,Agaev:2022ast,Braaten:2020nwp,Karliner:2021wju,
He:2022rta,Abreu:2022lfy,Agaev:2021vur,Azizi:2021aib,Vijande:2009kj,Navarra:2007yw,Yang:2009zzp,Tan:2020ldi,Achasov:2022onn,Lyu:2023xro,Li:2012ss,Cheng:2020wxa,Lin:2023gzm,Chen:2013aba,Wang:2020jgb}.
In the present study , it is generally accepted that $T_{cc}^{+}$ is a molecular state, a compact tetraquark or a triangle singularity.
For example, in Ref.~\cite{Du:2021zzh}, by utilizing coupled-channel approach, the authors suggested that $T_{cc}^{+}$ is a hadronic molecule generated by the interactions in the $D^{*+}D^{0}$ and $D^{*0}D^{+}$ channels. In Ref.~\cite{Huang:2021urd}, they also argued that $T_{cc}^{+}$ is molecule state in the framework of chiral unitary theory. Moreover, S. S. Agaev \textit{et al.}~\cite{Agaev:2021vur} proposed that $T_{cc}^{+}$ is a compact structure by employing QCD three-point sum rule method. In Refs.~\cite{Braaten:2022elw,Achasov:2022onn}, they advocated that this state may be triangle logarithmic singularity.

In addition to the $T_{cc}$ state, double-bottom ($T_{bb}$) tetraquark states have also been investigated in various theoretical frameworks. In Ref.~\cite{Li:2012ss}, utilizing a coupled-channel approach, the authors proposed that $\bar{B}^{*}\bar{B}^{*}$ states with $IJ^{P}=01^{+},11^{+}$ might be a good molecular candidates.
In Ref.~\cite{Cheng:2020wxa}, J. Cheng \textit{et al.} employed heavy diqaurk-antidiquark symmetry (HDAS) and the chromomagnetic interaction (CMI) model to study $bb\bar{n}\bar{n}$ ($n=u,~d$) tetraquark states. The results show that the lowest energy of $bb\bar{n}\bar{n}$ with $IJ^{P}=01^{+}$ is approximately $116$ MeV below the $\bar{B}\bar{B}^{*}$ threshold.
In Ref.~\cite{Lin:2023gzm}, Lin \textit{et al.} investigated the $P$-wave $T_{bb}^{-}$ states in the diqaurk-antidiquark picture by employing quark model. In the $\rho$-mode with $1^{1}P_{1}$ $[bb]_{\bar{3}}^{0}[\bar{u}\bar{d}]_{3}^{0}$ configuration, its energy is about $18$ MeV lower than the $\bar{B}\bar{B}$ threshold, which means it can establish a compact bound state.

Furthermore, theoretical investigations have explored other partners ($T_{bc}$, $T_{bc\bar{s}}$, $T_{cc\bar{s}}$, $T_{bb\bar{s}}$) of $T_{cc}$.
In the case of $T_{bc}$, $T_{bc\bar{s}}$ states, in Ref.~\cite{Deng:2021gnb}, by employing nonrelativistic quark model, the authors provided that $T_{bc}$ with $IJ^{P}=00^{+}$ and $01^{+}$, $T_{bcs}$ with $\frac{1}{2}0^{+}$, $\frac{1}{2}1^{+}$ can form a bound state. The high-spin states $T_{bc}$ with $02^{+}$ and $T_{bc\bar{s}}$ with $\frac{1}{2}2^{+}$ can decay into $D$-wave $\bar{B}D$ and $\bar{B}_{s}D$ although it is below the $\bar{B}^{*}D^{*}$ and $\bar{B}_{s}^{*}D^{*}$ thresholds, respectively.
QCD sum rule has also been used to calculate $T_{bc}$ and $T_{bc}$~\cite{Chen:2013aba,Wang:2020jgb}. In Ref.~\cite{Chen:2013aba}, the authors suggested that for the $T_{bc\bar{s}}$ states with $J^{P}=0^{+}$ and $1^{+}$, both masses are below the corresponding thresholds ($D^{(*)}\bar{B}^{(*)}$). In Ref.~\cite{Wang:2020jgb}, the results show that the masses of $T_{bc\bar{s}}$ tetraquark states are below the thresholds of $\bar{B}_{s}D$ and $\bar{B}^{*}_{s}D$ final states for the scalar and axial-vector channels, respectively.
Regarding $T_{cc\bar{s}}$ and $T_{bb\bar{s}}$ tetraquark states, in Ref.~\cite{Deng:2021gnb}, the authors argued that the $T_{cc\bar{s}}$ is not a bound state, but they obtained a shallow bound state for $T_{bb\bar{s}}$ tetraqaurk state with $IJ^{P}=\frac{1}{2}1^{+}$. In the heavy quark limit~\cite{Braaten:2020nwp}, via the color-coulomb potential in the diquark-antidiquark picture, the $T_{bb\bar{s}}$ state are bound into compact structures. However, in Ref.~\cite{Karliner:2021wju}, the results indicates that $T_{cc\bar{s}}$ can be establish stable state against the strong interaction.
Further results and discussions on the partners of $T_{cc}$ can be found in Refs.~\cite{Kim:2022mpa,Meng:2021yjr,Albuquerque:2023rrf,Agaev:2019lwh,Hernandez:2019eox,Carlson:1987hh,Pepin:1996id,
Pflaumer:2021ong,Guo:2021yws,Weng:2021hje,Dai:2021vgf}.

While $T_{cc}$ partners involving heavy quarks have been extensively studied, limited attention has been given to double strangeness partners ($T_{ss}$) tetraquark states. Thus, the investigation of bound states and resonant states in the $T_{ss}$ system is  both interesting and necessary.

The structure of the paper is as follows: Section II provides a brief description of the quark model and wave functions. Section III introduces the real-scaling method. Section IV details the $^3P_0$ method. Section V presents our results and discussion.
\section{Chiral quark model, wave function of $T_{ss}$ system} \label{wavefunction and chiral quark model}
\subsection{Chiral quark model}

In this paper, we primarily investigate the dynamics of the $s\bar{q}s\bar{q}$ system using the chiral quark model. This model is an excellent potential framework as it effectively explains hadron interactions, including mass terms, kinetic terms, and potential terms, which can be written as

\begin{eqnarray}
H &=& 2 (m_s + m_q) + 2 (\frac{\vec{p}_{s\bar{q}}^2}{2\mu_{s\bar{q}}}) +\frac{\vec{p}_{s\bar{q}s\bar{q}}^2}{2\mu_{s\bar{q}s\bar{q}}} + V(r_{ij}).
\end{eqnarray}
where  $m$ denotes the quark mass, $\mu$ represents the reduced mass, and $\vec{p}$ is the relative momentum between quarks. The form of $\mu$ is as follows,
\begin{eqnarray}
\mu_{s\bar{q}}&=&\frac{m_{s}  m_{\bar{q}}}{m_{s} + m_{\bar{q}}},\mu_{s\bar{q}s\bar{q}}=\frac{(m_{s}+m_{\bar{q}})}{2}.   \nonumber\\
\end{eqnarray}

In the $s\bar{q}s\bar{q}$ system, $s$ and $q$ can be approximately regarded as identical quarks. Consequently, the exchange of Goldstone bosons between $s$ and $q$ plays a significant role. Therefore, our potential includes confinement potential, one-gluon exchange potential, and Goldstone boson exchange potential, expressed as follows:

\begin{eqnarray}
V(r_{ij}) &=& \sum_{i<j=1}^n [ V_{con}(r_{ij})+V_{oge}(r_{ij}) + \sum_{\chi=\pi,\eta,K} V_{\chi}(r_{ij}) \nonumber\\
          &+& V_{\sigma}(r_{ij})]
\end{eqnarray}

Goldstone boson exchange mainly stems from the spontaneous chiral symmetry breaking, a fundamental characteristic of QCD. This results in a long-range attractive force between quarks. In our chiral quark model, the chiral part of the Hamiltonian, ($V_{\pi}(r_{ij})$+$V_{\eta}(r_{ij})$ +$V_{K}(r_{ij})$+ $V_{\sigma}(r_{ij})$), can be expressed as follows,
\begin{eqnarray}
V_{\pi}(r_{ij}) &=& \frac{g_{ch}^2}{4\pi}\frac{m_{\pi}^2}{12m_im_j}\frac{\Lambda_{\pi}^2}{\Lambda_{\pi}^2-m_{\pi}^2}m_\pi [ Y(m_\pi r_{ij})-           \nonumber  \\
&& \frac{\Lambda_{\pi}^3}{m_{\pi}^3}Y(\Lambda_{\pi} r_{ij}) ] \boldsymbol{\sigma}_i \cdot\boldsymbol{\sigma}_j \sum_{a=1}^3 \lambda_i^a \lambda_j^a, \nonumber  \\
V_{K}(r_{ij})&=& \frac{g_{ch}^2}{4\pi}\frac{m_{K}^2}{12m_im_j} \frac{\Lambda_K^2}{\Lambda_K^2-m_{K}^2}m_K   [ Y(m_K r_{ij})-                \nonumber  \\
&& \frac{\Lambda_{K}^3}{m_{K}^3}Y(\Lambda_{K} r_{ij}) ] \boldsymbol{\sigma}_i \cdot\boldsymbol{\sigma}_j \sum_{a=1}^3 \lambda_i^a \lambda_j^a,            \\
V_{\eta}(r_{ij})& = & \frac{g_{ch}^2}{4\pi}\frac{m_{\eta}^2}{12m_im_j}\frac{\Lambda_{\eta}^2}{\Lambda_{\eta}^2-m_{\eta}^2}m_{\eta} [ Y(m_\eta r_{ij})-  \nonumber  \\
&& \frac{\Lambda_{\eta}^3}{m_{\eta}^3}Y(\Lambda_{\eta} r_{ij}) ] \boldsymbol{\sigma}_i \cdot\boldsymbol{\sigma}_j [\lambda_i^8 \lambda_j^8 \cos\theta_P - \sin \theta_P],  \nonumber \\
V_{\sigma}(r_{ij})&=& -\frac{g_{ch}^2}{4\pi}\frac{\Lambda_{\sigma}^2}{\Lambda_{\sigma}^2-m_{\sigma}^2}m_\sigma [ Y(m_\sigma r_{ij})-\frac{\Lambda_{\sigma}}{m_\sigma}Y(\Lambda_{\sigma} r_{ij}) ],  \nonumber
\end{eqnarray}
where $\boldsymbol{\sigma}$ represents the Pauli matrices acting on the spin wave functions of the $T_{ss}$ system, while $\boldsymbol{\lambda}^{a}$ denotes the Gell-Mann matrices acting on the flavor wave functions of the $T_{ss}$ system. $Y(x)$ is the Yukawa function, explicitly given by $Y(x)=\frac{e^{-x}}{x}$. $\Lambda_{\chi}$ serves as the cut-off parameter, and $g^2_{ch}/4\pi$ denotes the Goldstone-quark coupling constant. The masses of Goldstone bosons are denoted by $m_{\pi}$, $m_{\eta}$, and $m_{K}$, while $m_{\sigma}$ is determined by the relation,
\begin{align}
m_{\sigma}^2 \approx m_{\pi}^2+4 m_{u,d}^2.
\end{align}

One of the characteristics of QCD is quark confinement, which suggests that quarks cannot be separated. This corresponds to the confinement potential in the Hamiltonian. Unlike other potential forms in the quark model, the confinement potential is not directly derived from field theory. Consequently, it typically appears in three common forms: linear confinement potential \cite{Yang:2017prf}, quadratic confinement potential \cite{Tan:2024omp}, and screened confinement potential \cite{Vijande:2004he}. These potentials share a common feature of exhibiting strong attraction at short distances. The distinction lies in the fact that due to the screening effects, screened confinement potential can describe well the  spectra of light mesons including excited states. Given that the $s\bar{q}s\bar{q}$ system involves light mesons ($K$/$K^{*}$), we employ the screened confinement potential in this study. Its specific form is given as follows:

\begin{align}
    V_{con} (r_{ij}) &=  [-a_{c}(1- e^{-\mu_c r_{ij}}) + \Delta ] \boldsymbol{\lambda}_i^c \cdot \boldsymbol{\lambda}_j^c,
\end{align}
where $\boldsymbol{\lambda}^{c}$ are $SU(3)$ color Gell-Mann matrices, $a_{c}$, $\mu_c$  are model parameters.

The one-gluon exchange potential $V_{oge}(r_{ij})$ reflects the QCD characteristic of asymptotic freedom, and plays a crucial role in the quark model by offering intermediate-range attractive forces. This potential generally consists of a coulomb term and a color-magnetic term. It reads

\begin{align}
 V_{oge} (r_{ij})&= \frac{\alpha_s}{4} \boldsymbol{\lambda}_i^c \cdot \boldsymbol{\lambda}_{j}^c
\left[\frac{1}{r_{ij}}-\frac{1}{6m_im_j}\boldsymbol{\sigma}_i\cdot
\boldsymbol{\sigma}_j   \frac{e^{-r_{ij}/r_0(\mu_{ij})}}{ r_{ij}r_0^2(\mu_{ij})}     \right],
\end{align}
where $\alpha_s$ is the QCD-inspired  strong coupling constant, which is is determined by fitting experimental meson data. Generally, the heavier the quark, the smaller the coupling constant between quarks. All the parameters are determined by fitting the meson spectrum, from light to heavy, taking into account only a
quark-antiquark component. They are shown in Table~\ref{modelparameters}.

\begin{table}[t]
\begin{center}
\caption{Quark model parameters ($m_{\pi}=0.7$ $fm^{-1}$, $m_{\sigma}=3.42$ $fm^{-1}$, $m_{\eta}=2.77$ $fm^{-1}$, $m_{K}=2.51$ $fm^{-1}$).\label{modelparameters}}
\begin{tabular}{cccc}
\hline\hline\noalign{\smallskip}
Quark masses   &$m_u=m_d$(MeV)     &313  \\
               &$m_{s}$(MeV)         &555  \\

Goldstone bosons   %&$m_{\pi}(fm^{-1})$     &0.70  \\
                   &$\Lambda_{\pi}=\Lambda_{\sigma}(fm^{-1})$     &4.2  \\
                   &$\Lambda_{\eta}=\Lambda_{K}(fm^{-1})$     &5.2  \\
                   &$g_{ch}^2/(4\pi)$                &0.54  \\
                   &$\theta_p(^\circ)$                &-15 \\

Confinement             &$a_{c}$ (MeV)     &430 \\
                        &$\mu_{c}$($fm^{-1}$)     &0.7 \\
                        &$\Delta$(MeV)       &181.1 \\

OGE                 & $\alpha_{qq}$        &0.54 \\
                    & $\alpha_{qs}$        &0.48 \\
                    & $\alpha_{ss}$        &0.42 \\
                    &$\hat{r}_0$(MeV)    &28.17 \\
\hline\hline
\end{tabular}
\end{center}
\end{table}

\subsection{The wave function of $T_{ss}$ system}
The $T_{ss}$ system has two important structures: the molecular configuration ($s\bar{q}$-$s\bar{q}$) and the diquark configuration ($ss$-$\bar{q}\bar{q}$). At the quark level, these are expressed as the product of four components: color, spin, flavor, and orbital parts. According to QCD, any quark system, including multi-quark states, must be colorless. Therefore, the color wave function of a four-quark system can be obtained by either two coupled color-singlet clusters, $\mathbf{1} \otimes \mathbf{1}$,

\begin{align}
C_{1} =  \sqrt{\frac{1}{3}}(\bar{r}r+\bar{g}g+\bar{b}b) \times \sqrt{\frac{1}{3}}(\bar{r}r+\bar{g}g+\bar{b}b).
\end{align}
Additionally, the color wave function of a colorless four-quark system can also be obtained by coupling two  color-octet  wav function, $\mathbf{8} \otimes \mathbf{8}$,
\begin{align}
C_{2} & =  \sqrt{\frac{1}{72}}(3\bar{b}r\bar{r}b+3\bar{g}r\bar{r}g+3\bar{b}g\bar{g}b+3\bar{g}b\bar{b}g+3\bar{r}g\bar{g}r \nonumber \\
&  +3\bar{r}b\bar{b}r+2\bar{r}r\bar{r}r+2\bar{g}g\bar{g}g+2\bar{b}b\bar{b}b-\bar{r}r\bar{g}g \nonumber \\
&  -\bar{g}g\bar{r}r-\bar{b}b\bar{g}g-\bar{b}b\bar{r}r-\bar{g}g\bar{b}b-\bar{r}r\bar{b}b). \nonumber \\
\end{align}
These two color wave functions correspond to the molecular configurations. In the case of the diquark structure, the color wave functions are constructed in a similar manner to the coupling of two color-octet states into a color-singlet four-quark wave function. Specifically, it involves coupling $\mathbf{3} \otimes \bar{\mathbf{3}}$ and $\mathbf{6} \otimes \bar{\mathbf{6}}$ to form a colorless four-quark system. The wave function can be expressed as follows,
\begin{eqnarray}
C_{3} &= &
 \sqrt{\frac{1}{12}}(rg\bar{r}\bar{g}-rg\bar{g}\bar{r}+gr\bar{g}\bar{r}-gr\bar{r}\bar{g}+rb\bar{r}\bar{b} \nonumber \\
 & & -rb\bar{b}\bar{r}+br\bar{b}\bar{r}-br\bar{r}\bar{b}+gb\bar{g}\bar{b}-gb\bar{b}\bar{g} \nonumber \\
 & & +bg\bar{b}\bar{g}-bg\bar{g}\bar{b}), \nonumber \\
C_{4} &= & \sqrt{\frac{1}{24}}(2rr\bar{r}\bar{r}+2gg\bar{g}\bar{g}+2bb\bar{b}\bar{b}
    +rg\bar{r}\bar{g}+rg\bar{g}\bar{r} \nonumber \\
& & +gr\bar{g}\bar{r}+gr\bar{r}\bar{g}+rb\bar{r}\bar{b}+rb\bar{b}\bar{r}+br\bar{b}\bar{r} \nonumber \\
& & +br\bar{r}\bar{b}+gb\bar{g}\bar{b}+gb\bar{b}\bar{g}+bg\bar{b}\bar{g}+bg\bar{g}\bar{b}).
\end{eqnarray}

In contrast to color wave functions, the spin of antiquarks and quarks is indistinguishable. Therefore, diquark and molecular structures share the same spin wave function.  Given that we are studying the $T_{ss}$ system with total spin $1$, its wavefunction exists in three possible configurations: $ \mathbf{0} \otimes \mathbf{1} $ ( $S_1$ ), $ \mathbf{1} \otimes \mathbf{0}$ ( $S_2$ ), and $\mathbf{1} \otimes \mathbf{1} $ ( $S_3$ ).
\begin{align*}
S_{1} &=\chi_{00}^{\sigma}\chi_{11}^{\sigma},\\
S_{2} &=\chi_{11}^{\sigma}\chi_{00}^{\sigma},\\
S_{3} &=\frac{1}{\sqrt{2}}(\chi_{11}^{\sigma}\chi_{10}^{\sigma}-\chi_{10}^{\sigma}\chi_{11}^{\sigma}).\\
\end{align*}

The flavor wave functions of the sub-clusters for the two structures are presented as follows:
\begin{eqnarray}
 && \chi_{\frac{1}{2},\frac{1}{2}}^{fm}=s\bar{d},\chi_{\frac{1}{2},-\frac{1}{2}}^{fm}=-s\bar{u},\\
 && \chi_{0,0}^{fd_{1}}=ss,\chi_{0,0}^{fd_{2}}= \frac{1}{\sqrt{2}}(\bar{u}\bar{d} - \bar{d}\bar{u}).
\end{eqnarray}
In this case, the subscripts of $\chi_{I,I_z}^{fm(d)i}$  denote the isospin and its third component, while the superscripts specify the structure ( $fm$ for the molecular configuration and
$fd$ for the diquark configuration ) and the index. Considering the isospin of $T_{cc}$, we assign the isospin of $T_{ss}$ to be zero. Thus, the total flavor wave function can be coupled in two possible ways:  $ \mathbf{\frac{1}{2}} \otimes \mathbf{\frac{1}{2}} $ ( molecular configuration, $F_1$) and $ \mathbf{0} \otimes \mathbf{0} $ ( diquark configuration, $F_2$). The total flavor wave functions can be written as,
\begin{eqnarray}
F_{1} & = & \frac{1}{\sqrt{2}}\left( \chi_{\frac{1}{2}\frac{1}{2}}^{fm}
 \chi_{\frac{1}{2}-\frac{1}{2}}^{fm}-\chi_{\frac{1}{2}-\frac{1}{2}}^{fm} \chi_{\frac{1}{2}\frac{1}{2}}^{fm} \right), \nonumber\\
F_{2} & = & \chi_{\frac{1}{2}\frac{1}{2}}^{fm}  \chi_{\frac{1}{2}\frac{1}{2}}^{fm}.
\end{eqnarray}

Since the $T_{ss}$ system has positive parity, we set all orbital quantum numbers to zero. Consequently, its spatial wavefunction can be represented as follows,
\begin{eqnarray}\label{spatialwavefunctions}
\psi(r) = \Psi_{l_1=0}({\bf r}_{12})\Psi_{l_2=0}({\bf r}_{34}) \Psi_{L_r=0}({\bf r}_{1234}). \nonumber
\end{eqnarray}

The spatial wave functions for the three relative motions involved here are expanded using the Gaussian Expansion Method (GEM).

\begin{align}
\Psi_{l_1=0}(r_{12}) & = \sum_{n1=1}^{n1_{\rm max}} c_{n1} N_{n1} e^{-\nu_{n1}r^2}Y_{00}(\hat{r_{12}}), \\
\Psi_{l_2=0}(r_{34}) & = \sum_{n2=1}^{n2_{\rm max}} c_{n2} N_{n2} e^{-\nu_{n2}r^2}Y_{00}(\hat{r_{34}}), \\
\Psi_{L_r=0}(r_{1234}) & = \sum_{n3=1}^{n3_{\rm max}} c_{n3} N_{n3} e^{-\nu_{n3}r^2}Y_{00}(\hat{r_{1234}}),
\end{align}
with the normalization constant,
\begin{equation}
N_{n}  = \left[\frac{4(2\nu_{n})^{\frac{3}{2}}}{\sqrt{\pi}} \right]^\frac{1}{2} . \\
\end{equation}
The coefficients $c_n$ are variational parameters determined through dynamic optimization. The Gaussian size parameters follow a geometric progression,
\begin{equation}\label{gaussiansize}
\nu_{n}=\frac{1}{r^2_n}, \quad r_n=r_1a^{n-1}, \quad
a=\left(\frac{r_{n_{\rm max}}}{r_1}\right)^{\frac{1}{n_{\rm max}-1}}.
\end{equation}
where $r_n$ denotes the Gaussian size parameter for the $n$-th term, with $r_1$ and $r_{n_{max}}$ as the initial parameters.

The total wave function is constructed as the direct product of the orbital, spin, color, and flavor wave functions. Thus, the complete wave function of the system is expressed
\begin{equation}
\Psi^{i,j,k}={\cal A} \psi(r) S_i F_j C_k.
\end{equation}
Here, $\cal A$ denotes the antisymmetrization operator. For the $s\bar{q}s\bar{q}$ system, where the two subsystems are identical, $\cal A$ is defined as $1-(13)-(24)+(13)(24)$. Subsequently, we solve the Schr\"{o}dinger equation below to determine the system's eigen-energies, utilizing the Rayleigh-Ritz variational principle
 \begin{equation}
H\Psi^{i,j,k}=E\Psi^{i,j,k}.
\end{equation}.

\section{Real-scaling method}

The real-scaling method was first proposed by Taylor \cite{Taylor:1970} to study resonance states in electron-molecule systems. Due to its success in explaining a substantial amount of experimental data, Hiyama \emph{et al.} \cite{Hiyama:2018ukv} applied this method at the quark level to search for the $P_c$ states in the $qqqc\bar{c}$ system. Subsequently, we introduced this method to the tetraquark system for the first time and successfully explained particles \cite{Tan:2019knr,Tan:2020cpu,Tan:2022pzi,Tan:2023azs,Wu:2023hhk} such as $Y(4626)$, $X(2900)$, and $\Upsilon(10753)$.

\begin{figure}[htp]
  \setlength {\abovecaptionskip} {-0.1cm}
  \centering
  \resizebox{0.50\textwidth}{!}{\includegraphics[width=3cm,height=2.2cm]{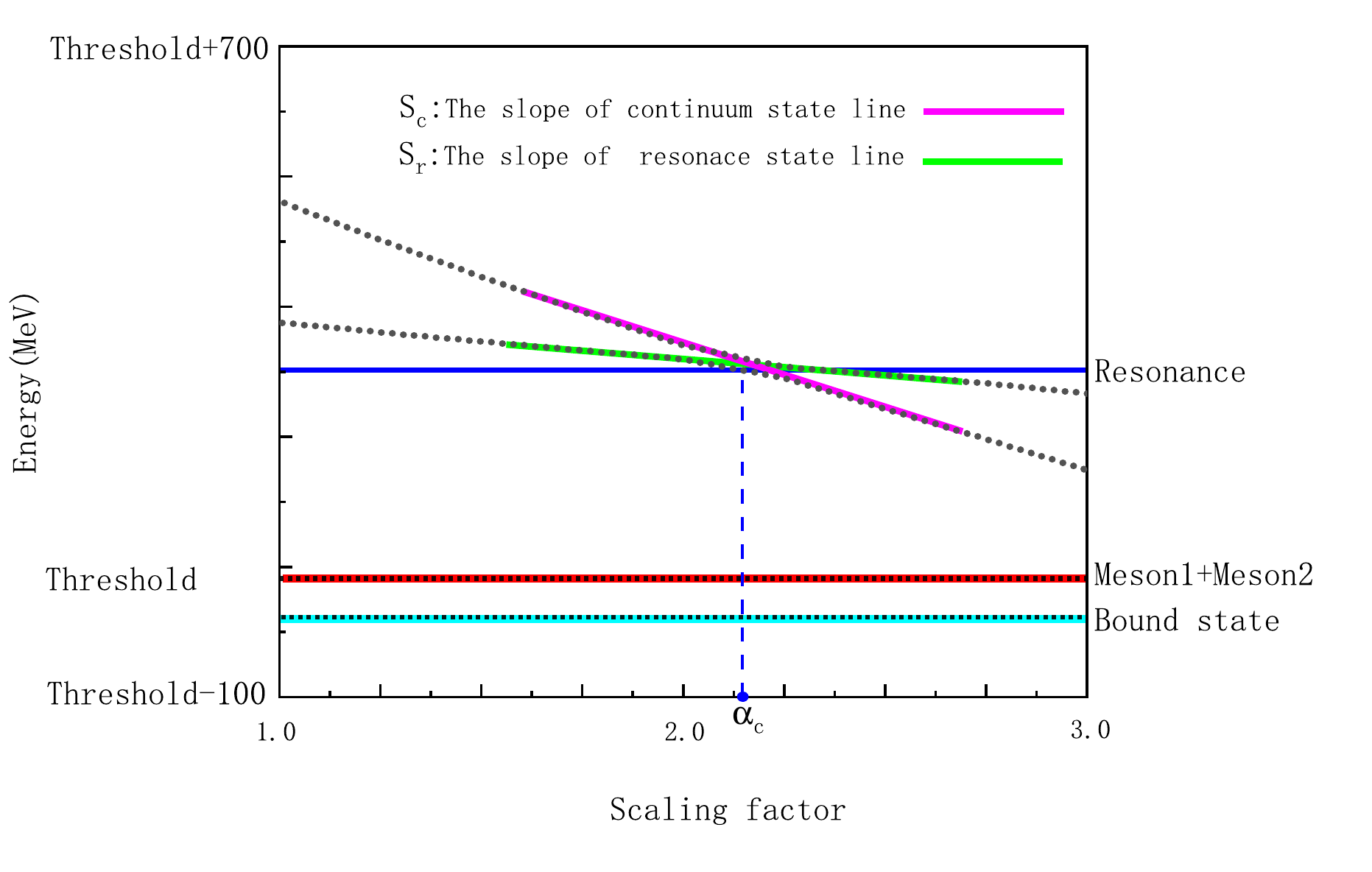}}
  \caption{Two forms of stable states, bound state (Horizontal line) and resonance state (avoiding-crossing structure), in the real-scaling method. The red line represents the threshold, while the light blue line indicates the bound state. The avoided-crossing structure formed by the red line, representing the scattering state, and the green line, representing the resonance, corresponds to a genuine resonance.}
\label{RSM}
\end{figure}

The real-scaling method involves systematically scaling the width of Gaussian functions between two groups using a scaling factor, denoted as $\alpha$, $R \rightarrow \alpha R$. As $\alpha$ increases, the width of the Gaussian functions expands, leading to variations in the energy of system. If a stable structure is present, it remains unaffected by the changes in Gaussian function widths. Within the framework of this method, bound states do not exhibit decay behavior due to the absence of corresponding decay channels, and thus manifest as horizontal lines, shown in FIG. \ref{RSM}(a). Conversely, resonance states, having several decay channels, strongly couple with scattering channels and decay to the relevant threshold channels. This is depicted in FIG. \ref{RSM}(b) by two downward-sloping lines that demonstrate avoided crossing behavior. Since resonance states are stable structures, the avoid-crossing structure may reappear as the scaling factor $\alpha$ increases. However, a larger $\alpha$ results in a more complex calculation process. Therefore, for computational efficiency, we chose the range of $\alpha$ to be between $1.0$ and $3.0$. Within this range, most resonance states will exhibit repeated avoid-crossing structures. The real-scaling method allows for the calculation of the decay width of stable states. The calculation equation is as follows,
\begin{equation}\label{formula_RSM}
\Gamma =4 |V(\alpha_c)|\frac{\sqrt{|Slope_r||Slope_s|}}{|Slope_r-Slope_s|}.
\end{equation}
Based on the discussion above, in the framework of real-scaling method, avoided-crossing structure represents resonant state, while horizontal line indicates bound state. Therefore, the decay width calculation Eq. \ref{formula_RSM} is applicable only to the decay width of resonant states. Here, $\alpha_c$ denotes  the $\alpha$ coordinate corresponding to the avoid-crossing structure, while $V(\alpha_c)$ represents the energy difference at the avoid-crossing point between the two lines. The  "$slope_r$" and "$slope_s$" indicate the slope of these two lines.

\section{$^3P_0$ model} \label{T operator}

The $^3P_0$ model is extensively employed in the calculation of meson decay widths \cite{Micu:1968mk,LeYaouanc:1972vsx}. It posits that a meson in the vacuum excites a pair of quark-antiquark, with momenta $\mathbf{p}_3$ and $\mathbf{p}_4$, respectively. The probability of their production is determined by a parameter $\gamma$, which is generally determined by an overall fitting of the strong decay width of hadrons. In this way, one obtains $\gamma=6.95$ for $u\bar{u}$ and $d\bar{d}$ pair creation, and $\gamma=6.95/\sqrt{3}$ for $s\bar{s}$ pair creation \cite{LeYaouanc:1977gm}. The flavor wave function of the quark-antiquark pair is given by $\phi_{0} = \frac{1}{\sqrt{3}}(u\bar{u} + d\bar{d} + s\bar{s})$, and the color wave function is $\omega_{0} = \frac{1}{\sqrt{3}}(r\bar{r} + g\bar{g} + b\bar{b})$. The operator in the model is,

\begin{eqnarray} \label{T0}
  &&T=-3~\gamma\sum_m\langle 1m1-m|00\rangle\int
  d\mathbf{p}_3d\mathbf{p}_4\delta^3(\mathbf{p}_3+\mathbf{p}_4)\nonumber\\
  &&~~~~\times{\cal{Y}}^m_1(\frac{\mathbf{p}_3-\mathbf{p}_4}{2})
  \chi^{34}_{1-m}\phi^{34}_0\omega^{34}_0b^\dagger_3(\mathbf{p}_3)d^\dagger_4(\mathbf{p}_4),
\end{eqnarray}

The S-matrix element for the process $A \rightarrow B + C$ is written as
 \begin{equation}
 \langle BC|T|A\rangle=\delta^3(\mathbf{P}_A-\mathbf{P}_B-\mathbf{P}_C){\cal{M}}^{M_{J_A}M_{J_B}M_{J_C}},
 \end{equation}
where $\mathbf{P}_B$ and $\mathbf{P}_C$ are the momenta of B and C mesons in the final state, and satisfy $\mathbf{P}_A = \mathbf{P}_B + \mathbf{P}_C = 0$ in the center-of-mass frame of meson A. ${\cal{M}}^{M_{J_A}M_{J_B}M_{J_C}}$ is the helicity amplitude of the process $A \rightarrow B + C$, which can be obtained as
 \begin{widetext}
 \begin{eqnarray}
 {\cal{M}}^{M_{J_A}M_{J_B}M_{J_C}}(\mathbf{P})&=&\gamma\sqrt{8E_AE_BE_C}
 \sum_{\renewcommand{\arraystretch}{.5}\begin{array}[t]{l}
 \scriptstyle M_{L_A},M_{S_A},\\\scriptstyle M_{L_B},M_{S_B},\\
 \scriptstyle M_{L_C},M_{S_C},m
 \end{array}}\renewcommand{\arraystretch}{1}\!\!
 \langle L_AM_{L_A}S_AM_{S_A}|J_AM_{J_A}\rangle
 \langle L_BM_{L_B}S_BM_{S_B}|J_BM_{J_B}\rangle\nonumber\\
 &&\times\langle
 L_CM_{L_C}S_CM_{S_C}|J_CM_{J_C}\rangle\langle 1m1-m|00\rangle
 \langle\chi^{14}_{S_BM_{S_B}}\chi^{32}_{S_CM_{S_C}}|\chi^{12}_{S_AM_{S_A}}\chi^{34}_{1-m}\rangle
 \nonumber\\
 &&\times[\langle \phi^{14}_B\phi^{32}_C|\phi^{12}_A\phi^{34}_0\rangle
 \mathcal{I}^{M_{L_A},m}_{M_{L_B},M_{L_C}}(\mathbf{P},m_1,m_2,m_3)\nonumber\\
 &&+(-1)^{1+S_A+S_B+S_C}\langle\phi^{32}_B\phi^{14}_C|\phi^{12}_A\phi^{34}_0\rangle
 \mathcal{I}^{M_{L_A},m}_{M_{L_B},M_{L_C}}(-\mathbf{P},m_2,m_1,m_3)],
 \end{eqnarray}
 with the momentum space integral
 \begin{eqnarray} \label{space integral}
 \mathcal{I}^{M_{L_A},m}_{M_{L_B},M_{L_C}}(\mathbf{P},m_1,m_2,m_3)=\int
 d\mathbf{p}\,\mbox{}\psi^\ast_{n_BL_BM_{L_B}}
 ({\scriptstyle\frac{m_3}{m_1+m_3}}\mathbf{P}+\mathbf{p})\psi^\ast_{n_CL_CM_{L_C}}
 ({\scriptstyle\frac{m_3}{m_2+m_3}}\mathbf{P}+\mathbf{p})
 \psi_{n_AL_AM_{L_A}}
 (\mathbf{P}+\mathbf{p}){\cal{Y}}^m_1(\mathbf{p}),\label{space}
 \end{eqnarray}
 \end{widetext}
 where $\mathbf{P}=\mathbf{P}_B=-\mathbf{P}_C$, and $\mathbf{p}=\mathbf{p}_3$, $m_3$ is the mass of the created quark $q_3$. To analyze the results and to compare the theoretical results with experimental data, the partial wave amplitude ${\mathcal{M}}^{JL}(A\rightarrow BC)$ is often employed. It is related with the helicity amplitude by the Jacob-Wick formula~\cite{Jacob:1959at},
 \begin{eqnarray}
 &&{\mathcal{M}}^{J L}(A\rightarrow BC) = \frac{\sqrt{2 L+1}}{2 J_A
 +1} \!\! \sum_{M_{J_B},M_{J_C}} \langle L 0 J M_{J_A}|J_A
 M_{J_A}\rangle \nonumber\\&&\times\langle J_B M_{J_B}
 J_C M_{J_C} | J M_{J_A} \rangle \mathcal{M}^{M_{J_A} M_{J_B}
 M_{J_C}}({\textbf{P}}). \label{MJL}
 \end{eqnarray}

Finally, the Eq. \ref{partialwidth} for calculating the decay width is as follows,
\begin{equation}
\Gamma = \pi^2 \frac{{|\textbf{P}|}}{M_A^2}\sum_{JL}\Big
|\mathcal{M}^{J L}\Big|^2, \label{partialwidth}
\end{equation}
where nonrelativistic phase-space is assumed,
\begin{equation}
|\textbf{P}| =\frac{\sqrt{[M^2_A-(M_B+M_C)^2][M^2_A-(M_B-M_C)^2]}}{2M_A},
\end{equation}
with $M_A$, $M_B$, $M_C$ being the masses of the mesons involved.  In this work, for calculational convenience, we transformed the transition operator $T$ from momentum space to coordinate space. The specific form is as follows,
\begin{eqnarray} \label{T2}
   T_2&=&-3\gamma\sum_{m}\langle 1m1-m|00\rangle\int
   d\mathbf{r_3}d\mathbf{r_4}(\frac{1}{2\pi})^{\frac{3}{2}}ir2^{-\frac{5}{2}}f^{-5}
   \nonumber \\
   &&Y_{1m}(\hat{\mathbf{r}})e^{-\frac{r^2}{4f^2}}e^{-\frac{R_{AV}^2}{f_0^2}}\chi_{1-m}^{34}\phi_{0}^{34}
   \omega_{0}^{34}b_3^{\dagger}(\mathbf{r_3})d_4^{\dagger}(\mathbf{r_4}).
\end{eqnarray}
Here, $\mathbf{R}_{AV}=\mathbf{R}_A-\mathbf{R}_V$ is the relative coordinate between the source particle
``A" and the created quark-antiquark pair in the vacuum with
\begin{eqnarray}
   \mathbf{R}_A & = & \frac{m_1\mathbf{r_1}+m_2\mathbf{r_2}}{m_1+m_2}; \nonumber \\
   \mathbf{R}_V & = & \frac{m_3\mathbf{r_3}+m_4\mathbf{r_4}}{m_3+m_4}=\frac{\mathbf{r_3}+\mathbf{r_4}}{2}.
   \nonumber
\end{eqnarray}

The convergence factor $e^{-r^2/(4f^2)}$ in the modified operator $T_2$ accounts primarily for the energy considerations of the quark-antiquark pairs created in the vacuum, acknowledging the difficulty of creating high-energy quark-antiquark pairs. The damping factor $e^{-R_{AV}^2/{R_0}^2}$ reflects that the created quark-antiquark pair should not be too far from the originating particle. Finnally, the parameter $\gamma = 6.95$ was initially obtained by fitting the heavy-flavor meson $c\bar{c}$. However, our previous work demonstrated that this parameter cannot be directly applied to the calculation of mass shifts in the light meson spectrum \cite{Chen:2017mug}. In this context, we need to calculate the decay of the light meson $K^{*}$, and therefore, we first obtained a new value for $\gamma$ by fitting the experimental data for $\rho \rightarrow \pi\pi$.  The values of the parameters $R_0$ and $f$ were ultimately determined by fitting the masses of light mesons,

\begin{eqnarray*}
\gamma = 32.2, \quad f = 0.5~ \mathrm{fm}, \quad R_0 = 1.0~ \mathrm{fm}.
\end{eqnarray*}

\section{Results and discussions}

In this section, we systematically studied the bound and resonant states in $T_{ss}$.  We coupled the diquark and molecular structures to explore potential bound states, $KK^{*}$. Within the framework of the four-quark model, the $KK^{*}$ state cannot decay via strong interactions into other two-meson channels. Therefore, the decay width of the bound state $KK^{*}$ can only originate from the process of $K^{*} \rightarrow K \pi$. Using the $^3P_0$ model, we first calculated the decay width of the $K^{*}$ in vacuum based on previously fitted parameters of light mesons. Then, we calculated the decay width of the $K^{*}$ when bound in the $KK^{*}$ state, taking into account the effects of binding energy. By comparing these two result, we estimated the decay width of the bound state $KK^{*}$. Finally, We then employed the real-scaling method to systematically identify resonant states in the energy range of 1.4 GeV to 2.8 GeV and calculated their decay widths.

\begin{table}[!t]
\caption{\label{Boundstate} Results of the bound state calculations in the $s\bar{q}s\bar{q}$ system.(unit: MeV)}
\begin{ruledtabular}
\begin{tabular}{ccccc}
Channel                                 & $|S_i F_j C_k\rangle(\Psi^{i,j,k})$  & E       & Mixed~~ \\
 $ KK^{*}$                              & $|111\rangle$          & $1393$  \\
 $ [K]_8[K^{*}]_8$                      & $|112\rangle$          & $1959$  \\
 $K^{*}K^{*}$                           & $|311\rangle$          & $1816$  \\
 $[K^{*}]_8[K^{*}]_8$                   & $|312\rangle$          & $1797$  \\
 \multicolumn{3}{c}{ coupled-molecular-channels:}  & $1392$   \\
 $[ss]_6^0[\bar{q}\bar{q}]_{\bar{6}}^1$ & $|124\rangle$          & $1974$  \\
 $[ss]_3^1[\bar{q}\bar{q}]_{\bar{3}}^0$ & $|223\rangle$          & $1543$  \\
 \multicolumn{3}{c}{ coupled-diquark-channels:}  & $1484$   \\
 \multicolumn{3}{c}{ complete coupled-channels:}  & $1328$   \\
\multicolumn{3}{c}{ Threshold($ K$+$K^{*}$):}  & $1392$   \\
\end{tabular}
\end{ruledtabular}
\end{table}
\subsection{Bound-state calculation:}

TABLE \ref{Boundstate} exhibits the results of the bound-state calculations. Due to symmetry restrictions, the molecular structure of the $T_{ss}$ system with $01^{+}$ involves four physical channels: two color singlets, $KK^{*}$ and $K^{*}K^{*}$, and their corresponding color octets. Meanwhile, the diquark structure of the $T_{ss}$ system contains only two physical channels: $[ss]_6^0[\bar{q}\bar{q}]_{\bar{6}}^1$ and $[ss]_3^1[\bar{q}\bar{q}]_{\bar{3}}^0$.

The energy of $KK^{*}$ is $1393$ MeV, which is 1 MeV above the threshold ($1392$ MeV). Its color octet state, $[K]_8[K^{*}]_8$, has an energy of $1959$ MeV, which is significantly higher than the threshold energy, indicating that neither of these channels form bound states. On the other hand, $K^{*}K^{*}$ has an energy of $1816$ MeV, whereas its color octet state, $[K^{*}]_8[K^{*}]_8$, has an energy of $1797$ MeV. This energy is not only lower than that of $[K]_8[K^{*}]_8$, but also lower than the energy of $K^{*}K^{*}$. This can be attributed to the fact that, within the quark model, strong attractive forces are more likely to occur between double vector mesons \cite{Wu:2021ahn}. We performed channel coupling for these four molecular states. The results indicate that, although $K^{*}K^{*}$ and $[K^{*}]_8[K^{*}]_8$ exhibit strong attractive interactions, the coupling effect still do not lower the energy of the lowest energy channel $KK^{*}$ below the threshold. Similarly, due to symmetry constraints, the diquark structure contains only two physical channels, $[ss]_6^0[\bar{q}\bar{q}]_{\bar{6}}^1$ and $[ss]_3^1[\bar{q}\bar{q}]_{\bar{3}}^0$. Among these, the $3 \times \bar{3}$ configuration has a relatively low energy of $1543$ MeV, just over 100 MeV above the threshold energy, whereas the $6 \times \bar{6}$ configuration has a much higher energy, heavier that of all the molecular structures. This suggests that the  diquark structure with $3 \times \bar{3}$ exhibits stronger attraction. Subsequently, we performed structural mixing of the two physical channels in the diquark structure and the four physical channels in the molecular structure. The calculation results indicate that we obtained a bound state, $B(1320)$, with a binding energy of 60 MeV. The composition analysis reveals that the molecular state $KK^{*}$ contributes $82\%$ to $B(1320)$, while the $[ss]_3^1[\bar{q}\bar{q}]_{\bar{3}}^0$ state contributes $13\%$, shown in Table \ref{Boundstate2}. The root-mean-square distance of $B(1320)$ shows that the distance between internal quarks is around $1$ fm, directly confirming that $B(1320)$ is predominantly a molecular state.

\begin{table}[!t]
\caption{\label{Boundstate2} The main components of the bound state and the root-mean-square distance among the internal quarks.(unit: fm)}
\begin{ruledtabular}
\begin{tabular}{cccccccccc}
   state     & $KK^{*}$  & $[ss]_3^1[\bar{q}\bar{q}]_{\bar{3}}^0$ &  $K^{*}K^{*}$ &$[K^{*}]_8[K^{*}]_8$  & $r_{s\bar{s}}$& $r_{s\bar{q}}$& $r_{\bar{q}\bar{q}}$\\
   $B(1320)$ & $82\%$    &  $13\%$                                &    $3\%$      & $2\%$                &   1.0  &   1.1 &   1.3 \\
\end{tabular}
\end{ruledtabular}
\end{table}

 Similar to the decay width of $T_{cc}$, which mainly comes from $DD^{*} \rightarrow DD + \pi$, the decay width of the predicted bound state $B(1320)$ should mainly result from $KK^{*} \rightarrow KK + \pi$. $T_{cc}$  has a binding energy of just $1$ MeV and hence only minimally affects the phase space for $DD^{*} \rightarrow DD + \pi$, resulting in a decay width almost equal to that of $D^{*}$. In contrast to $T_{cc}$, $B(1320)$ has a binding energy of 60 MeV, which substantially reduces the phase space for $KK^{*} \rightarrow KK + \pi$. As a result, the decay width of $B(1320)$ is expected to be less than that of $K^{*}$. To estimate the decay width of the bound state in the $T_{ss}$ system ($KK^{*}$), we distribute the binding energy of $60$ MeV equally between the two particles $K$ and $K^{*}$. Consequently, the energy of the $K^{*}$ in the $KK^{*}$ needs to be corrected by reducing it by approximately $30$ MeV. The decay width of the $K^{*}$, after this mass correction, provides the decay width of the bound state in the $T_{ss}$ system ($KK^{*}$). Employing the $^3P_0$ model, we calculated the decay width of $K^{*}$ and the decay width of $K^{*}$ within the bound state $KK^{*}$ separately. The results are listed in Table \ref{Boundstate3}. We observed that due to the influence of the binding energy, the decay width of the $K^{*}$ decreased from $20$ MeV to $17$ MeV. Therefore, we believe that the width of the bound state $B(1320)$ we obtained should be about 3 MeV smaller than the experimentally observed width of $K^{*}$.

\begin{table}[!t]
\caption{\label{Boundstate3} Comparison of the calculated decay widths of $K^{*}$ in vacuum and in the $KK^{*}$ bound State.}
\begin{ruledtabular}
\begin{tabular}{cccccccccc}
   state     & $K\pi$   & state     & $K\pi$ \\
   $K^{*}$       & $20$ MeV & $K^{*}$ in the $KK^{*}$       & $17$ MeV    \\
\end{tabular}
\end{ruledtabular}
\end{table}

\subsection{Resonant-state  calculation:}

In the quark model, resonances can form through two kinds of mechanism: the first involves strong attraction between mesons, resulting in resonance states, and the second involves resonance formation due to attractive interactions between colorful structures. Since in the $T_{ss}$ bound state calculations, every single channel is a scattering state, the first mechanism does not apply. Thus, we focus on identifying resonances under the second mechanism framework.

\begin{table}[!t]
\caption{\label{Resonantstate} Candidates for resonant states in the $s\bar{q}s\bar{q}$ system. (unit: MeV)}
\begin{ruledtabular}
\begin{tabular}{ccccc}
  ~~Resonances~~& $[ss]_6^0[\bar{q}\bar{q}]_{\bar{6}}^1$ &$[ss]_3^1[\bar{q}\bar{q}]_{\bar{3}}^0$ & $[K]_8[K^{*}]_8$   & $[K^{*}]_8[K^{*}]_8$\\
 ~~$E(1484)$~~  &    8.7\% ~~ &   83.3\%  ~~& ~1.1\% ~~ & ~~6.9\%  ~~\\
 ~~$E(1875)$~~  &   68.2\%~~  & ~~ 1.0\%  ~~& 12.8\% ~~ &  18.0\%~~\\
 ~~$E(2099)$~~  & ~~12.0\%~~  & ~~63.4\%  ~~&  5.6\% ~~ &  18.9\%~~\\
 ~~$E(2229)$~~  & ~~ 1.3\%~~  & ~~69.0\%  ~~& 18.6\% ~~ &  11.1\%~~\\
 ~~$E(2313)$~~  & ~~68.3\%~~  & ~~ 9.2\%  ~~&  8.4\% ~~ &  14.1\%~~\\
 ~~$E(2345)$~~  & ~~ 0.9\%~~  & ~~59.6\%  ~~& 21.1\% ~~ &  18.4\%~~\\
 ~~$E(2541)$~~  & ~~17.0\%~~  & ~~56.0\%  ~~& 15.3\% ~~ &  11.7\%~~\\
 ~~$E(2618)$~~  & ~~34.9\%~~  & ~~ 9.9\%  ~~& 31.5\% ~~ &  23.6\%~~\\
 ~~$E(2631)$~~  & ~~ 2.9\%~~  & ~~59.3\%  ~~& 25.5\% ~~ &  12.1\%~~\\
 ~~$E(2669)$~~  & ~~18.4\%~~  & ~~33.0\%  ~~&  8.2\% ~~ &  40.4\%~~\\
 ~~$E(2681)$~~  & ~~4.3\%~~  & ~~ 49.1\%  ~~& 42.2\% ~~ &  6.3\%~~\\
 ~~$E(2726)$~~  & ~~45.5\%~~  & ~~19.2\%  ~~& 11.4\% ~~ &  23.9\%~~\\
 ~~$E(2751)$~~  & ~~1.4\%~~  & ~~ 60.7\%  ~~& 18.3\% ~~ &  19.5\%~~\\
\end{tabular}
\end{ruledtabular}
\end{table}

\begin{figure*}[htbp]
\centering

\subfigure[Energy range from 1.2 GeV to 2.2 GeV]{
\begin{minipage}[t]{0.5\linewidth}
\centering
\includegraphics[width=9cm,height=11cm]{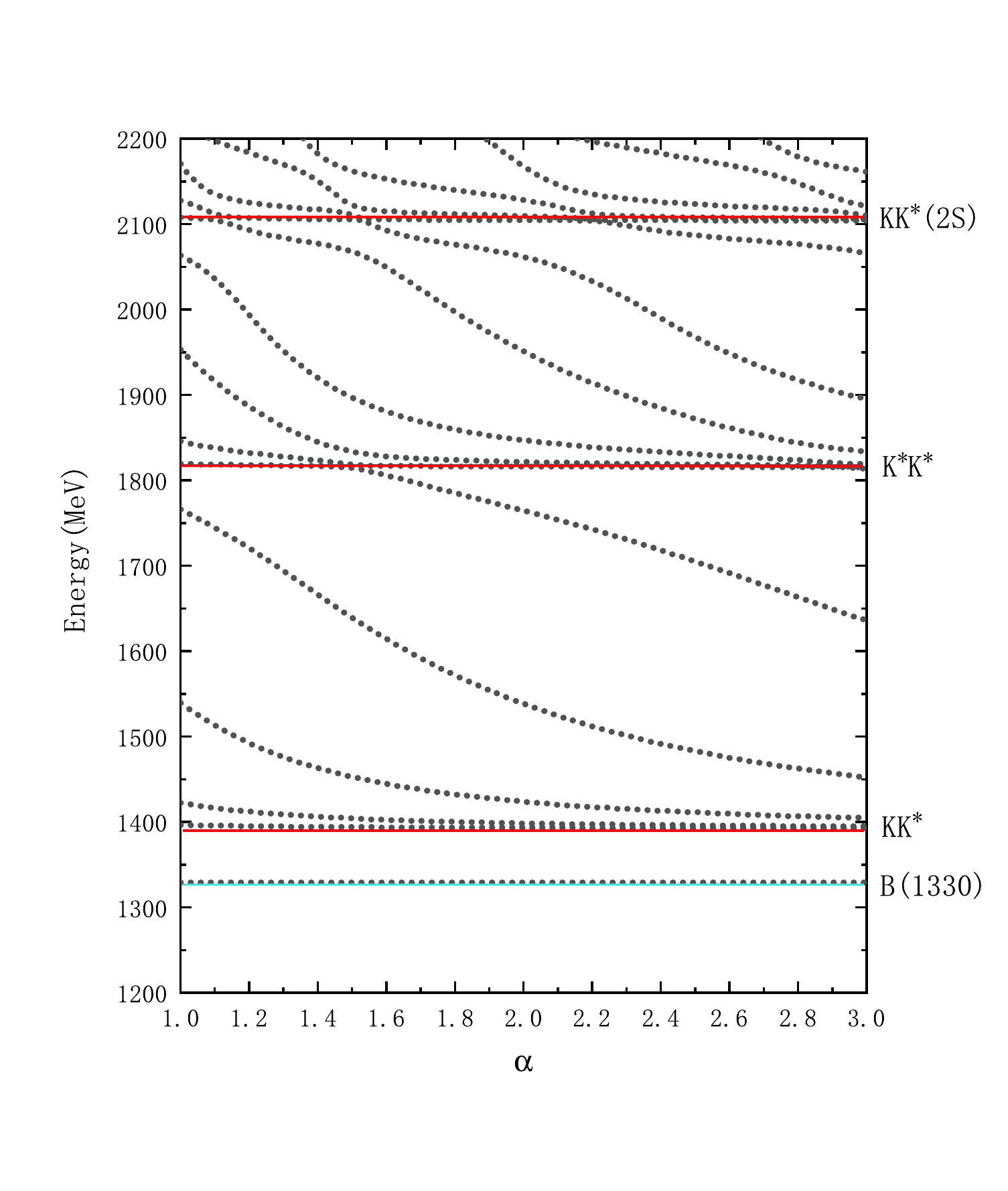}
\end{minipage}%
}%
\subfigure[Energy range from 2.2 GeV to 2.8 GeV]{
\begin{minipage}[t]{0.5\linewidth}
\centering
\includegraphics[width=9cm,height=11cm]{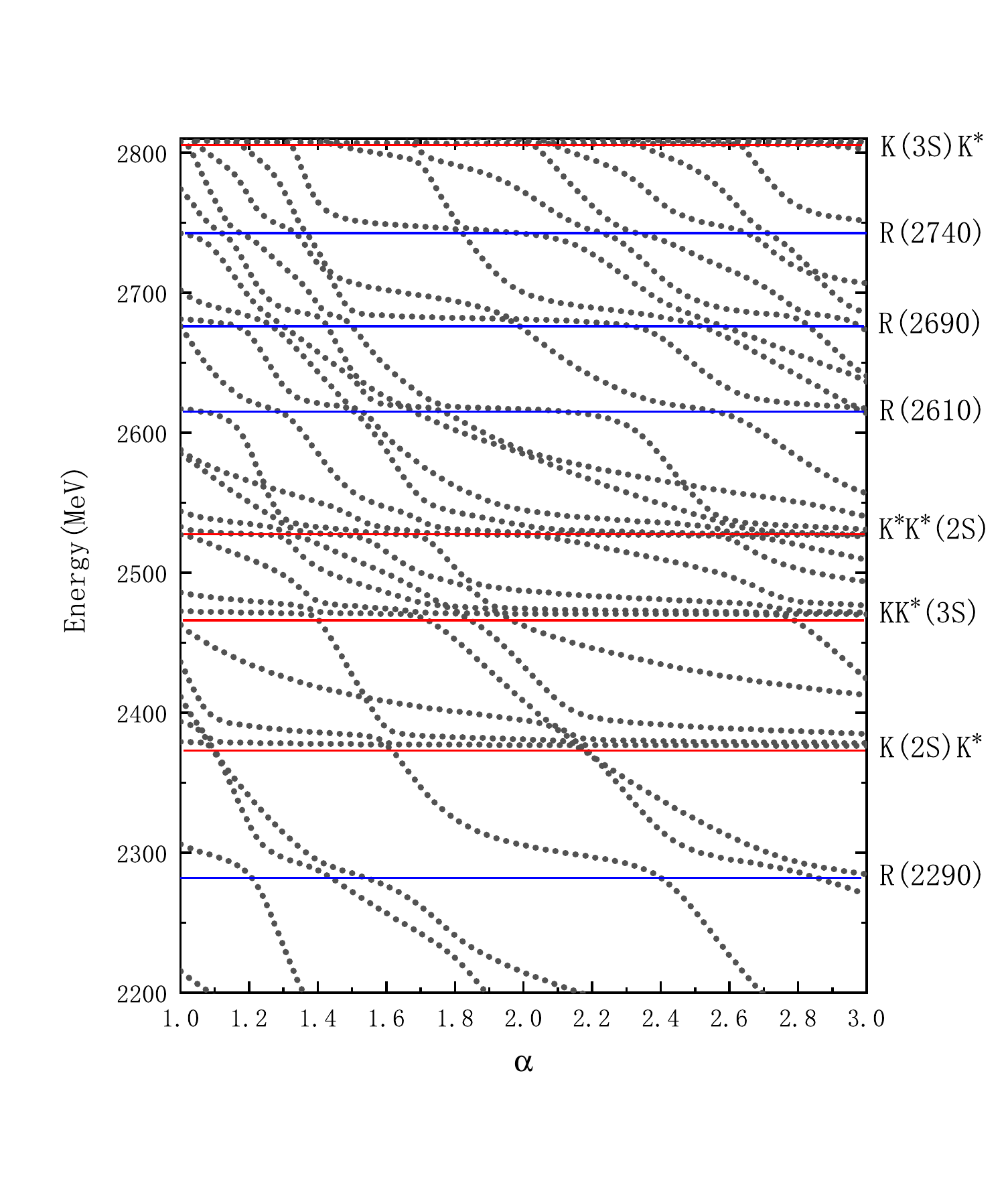}
\end{minipage}%
}

\centering
\caption{ Energy spectrum of $T_{ss}$ system by real-scaling method.}\label{Resonance2}
\end{figure*}

\begin{table*}[htb]
\caption{\label{Resonances3}Decay widths of the resonance states, main components of the resonance states (unit: MeV), and root-mean-square distances (unit: fm) in $ss\bar{q}\bar{q}$ system.}
\begin{tabular}{ccccccccccccccccccccc}\hline\hline
    state&   ~~   width&~~~$KK^{*}$&$[K]_8[K^{*}]_8$&$K^{*}K^{*}$&$[K^{*}]_8[K^{*}]_8$&$[ss]_3^1[\bar{q}\bar{q}]_{\bar{3}}^0$&$[ss]_6^0[\bar{q}\bar{q}]_{\bar{6}}^1$& ~~~$r_{s\bar{s}}$~~~&~~~ $r_{s\bar{q}}$~~~& ~~~$r_{\bar{q}\bar{q}}$~~~\\
$R(2290)$&~~~10.2&~~~$17.2\%$ ~~~&$8.0\%$ ~~~~ &$9.8\%$  ~~~~&$36.4\%$ ~~~~&$5.8\%$  ~~~~&$22.4\%$~~~~&~~~    1.1      ~~~ &~~~      1.2~~~     & ~~~   1.4 ~~~ \\
$R(2620)$&~~~5.3 &~~~$19.6\%$ ~~~&$24.8\%$ ~~~~&$0.9\%$  ~~~~&$36.3\%$ ~~~~&$15.0\%$ ~~~~&$3.2\%$ ~~~~& ~~~   1.2      ~~~ &~~~      1.4~~~     & ~~~   1.5 ~~~ \\
$R(2680)$&~~~5.8 &~~~$5.4\%$  ~~~&$25.5\%$ ~~~~&$15.1\%$ ~~~~&$28.9\%$ ~~~~&$7.3\%$  ~~~~&$17.6\%$~~~~& ~~~   1.2      ~~~ &~~~      1.3~~~     & ~~~   1.4 ~~~ \\
$R(2740)$&~~~1.9 &~~~$7.4\%$  ~~~&$14.5\%$ ~~~~&$1.6\%$  ~~~~&$35.1\%$ ~~~~&$0.3\%$  ~~~~&$40.1\%$~~~~& ~~~   0.9      ~~~ &~~~      1.0~~~     & ~~~   1.2 ~~~ \\
\hline
\end{tabular}
\end{table*}

In the $T_{ss}$ system, there are four colorful structures: two diquark configurations, $[ss]_6^0[\bar{q}\bar{q}]_{\bar{6}}^1$ and $[ss]_3^1[\bar{q}\bar{q}]_{\bar{3}}^0$, and two color-octet states, $[K]_8[K^{*}]_8$ and $[K^{*}]_8[K^{*}]_8$. We first performed channel coupling of these four colorful structures.  According to our calculations, within the energy range of $1.4$-$2.8$ GeV, we identified 13 possible  candidates of resonance in $T_{ss}$ system. These resonances, denoted as $E(\text{energy})$, are listed in Table \ref{Resonantstate} along with their corresponding color structure compositions. The majority of the candidates are situated around 2.6 GeV, indicating several genuine resonances at this energy level. Subsequently, we employed the real-scaling method to evaluate the stability of these resonances, with the results presented in FIG. \ref{Resonance2}. As shown in  FIG. \ref{Resonance2}, most of the resonance candidates decayed into their corresponding threshold channels. Only four resonances, $R(2290)$, $R(2610)$, $R(2690)$, and $R(2740)$, survived the coupling process with the scattering channels. The resonance $R(2290)$ primarily originates from the previous resonance candidate $E(2313)$. As the scaling factor $\alpha$ increased, the calculational  space changed, causing the energy of $E(2313)$ dropping to  $2290$ MeV, where it stabilized and formed an avoided crossing structure at $\alpha=1.2$.  Since the energies of most resonance candidates are concentrated around 2.6 GeV, we observe that the other three genuine resonances that survive after coupling with the scattering channels also have energies near 2.6 GeV. By comparing the energy levels, we can infer that resonance $R(2620)$ likely originates from candidate $E(2618)$, resonance $R(2680)$ from candidate $E(2682)$, and resonance $R(2740)$ from candidate $E(2751)$. We also calculated the percentage contributions of each channel and the root-mean-square  distance between quarks for these resonances, as listed in Table \ref{Resonances3}. The table reveals that each resonance has a significant $[K^{*}]_8[K^{*}]_8$ component. From our previous bound state calculations, we have already analyzed that the energy of $[K^{*}]_8[K^{*}]_8$ is lower than that of $K^{*}K^{*}$, indicating a strong attraction between $[K^{*}]_8[K^{*}]_8$. Since these four resonances contain substantial colorful structure components, their internal quark distances are small, within 1 fm. Additionally,  their decay widths are all less than $10$ MeV.

\section{summary}\label{Summary}
 In the framework of the chiral quark model, we systematically investigated the bound states and resonances of the $T_{ss}$ ($s\bar{q}s\bar{q}$ with $01^{+}$) system. Using the method of Gaussian expansion, we considered four physical channels, $KK^{*}$ and $K^{*}K^{*}$, and their corresponding color octets $[K]_8[K^{*}]_8$ and $[K^{*}]_8[K^{*}]_8$, in the molecular structure and two physical channels,$[ss]_6^0[\bar{q}\bar{q}]_{\bar{6}}^1$ and $[ss]_3^1[\bar{q}\bar{q}]_{\bar{3}}^0$, in the diquark structure, and performed a coupled-channel calculation of these two structures.

 The bound state calculations reveal a deeply bound state, $B(1330)$, with $KK^{*}$ as its primary component. We propose that the main decay width of this bound state arises from the decay process $K^{*} \rightarrow K + \pi$. Thus, using the $^3P_0$ model, we calculated the decay width of $K^{*}$ as $20$ MeV. However, the binding energy of $B(1330)$, approximately $60$ MeV, directly affects the decay phase space of $K^{*}$. By distributing this binding energy equally between $K^{*}$ and $K$, we estimate that the energy of $K^{*}$ should be reduced by about $30$ MeV during decay process. Consequently, the decay width of the bound state $B(1330)$ in the $T_{ss}$ system is estimated to be $17$ MeV. Subsequently, we employed the real-scaling method to calculate the resonances. The results indicate that we obtained four resonances: $R(2290)$, $R(2610)$, $R(2690)$, and $R(2740)$, primarily composed of colorul structure components. Their decay widths are all within $10$ MeV, and the internal quark distances are around $1$ fm.

 Given the current interest in $T_{cc}$, with much attention focused on its heavy-flavor partner, there is limited research on the light-flavor partner of $T_{cc}$. Therefore, we suggest that related experiments be conducted to search for the predicted bound states and resonances in $T_{ss}$ system.

\acknowledgments{This work is supported partly by the National Science Foundation of China under Contract No. 12205249 and the Funding for School-Level Research Projects of Yancheng Institute of Technology (No. xjr2022039, xjr2022040, xjr2020038).

\bibliographystyle{unsrt}

\end{document}